\documentclass{he_symp}
\usepackage{psfig,graphicx,epsfig}
\usepackage{color}
\usepackage{amsmath,amssymb,epic,eepic,array}
\begin{document}
\renewcommand{\FirstPageOfPaper }{ 206}\renewcommand{\LastPageOfPaper }{ 208}
\def\la{\mathrel{\mathchoice {\vcenter{\offinterlineskip\halign{\hfil
$\displaystyle##$\hfil\cr<\cr\sim\cr}}}
{\vcenter{\offinterlineskip\halign{\hfil$\textstyle##$\hfil\cr
<\cr\sim\cr}}}
{\vcenter{\offinterlineskip\halign{\hfil$\scriptstyle##$\hfil\cr
<\cr\sim\cr}}}
{\vcenter{\offinterlineskip\halign{\hfil$\scriptscriptstyle##$\hfil\cr
<\cr\sim\cr}}}}}
\def\ga{\mathrel{\mathchoice {\vcenter{\offinterlineskip\halign{\hfil
$\displaystyle##$\hfil\cr>\cr\sim\cr}}}
{\vcenter{\offinterlineskip\halign{\hfil$\textstyle##$\hfil\cr
>\cr\sim\cr}}}
{\vcenter{\offinterlineskip\halign{\hfil$\scriptstyle##$\hfil\cr
>\cr\sim\cr}}}
{\vcenter{\offinterlineskip\halign{\hfil$\scriptscriptstyle##$\hfil\cr
>\cr\sim\cr}}}}}



\title{Towards a consistent model for Neutron-Star Sources}
\author{W.Kundt}  
\institute{Institut f\"ur Astrophysik der Universit\"at,
Auf dem H\"ugel 71, 53121 Bonn, Germany}
\maketitle

\begin{abstract}
We are still far from understanding how pulsars pulse, how neutron stars are born, 
what neutron stars can emit, and in which way they do this. In this short communication, 
I list 18 alternatives -- several of them old, a few of them new -- which are handled 
differently by different authors but all 
of which are crucial for a model of neutron stars to be viable.
\end{abstract}

\section{Ten Alternatives related to Pulsars}  

$\bullet$ (1) We speak of pulsars as of `radio pulsars' even though the spectra $\nu S_{\nu} \ vs \ \nu$ 
of at least seven, perhaps already sixteen \emph{$\gamma$-ray pulsars} peak at hard $\gamma$-rays, between 
$\la$ MeV and $\ga$ 30 GeV photon energies. Uncertain in the spectra 
is their beaming fraction f = f($\nu$), which may well be frequency dependent. Is it justified to 
estimate the emitted power L via L := S $d^2$, rather than via L = 4 $\pi$ S $d^2$ ? The factor 4 $\pi$ 
matters in particular when we compare the bolometric luminosity of a pulsar with its 
spin-down power: What fraction of the spin-down power escapes as radiation, and what fraction in the wind? 

For a clarification, on the observational side we need a larger number of broadband spectra; and on 
the theoretical side we need an insight into how and at what \emph{separations} these spectra are emitted. 
Ever since Kundt \& Krotscheck (1980), it is my reinforced conviction that the (coherent) 
radio pulses are emitted somewhere inside the magnetosphere (as is widely agreed), logarithmically 
halfway between the stellar radius R and the speed-of-light distance c/$\Omega$, but that the much larger 
power at $\gamma$-rays (and X-rays) relies on a (further) transfer of 
spin-down power, downstream beyond c/$\Omega$, from the co-rotating magnetosphere to the outgoing, 
radiating leptons, i.e. that the $\gamma$-rays are emitted in the ``rural areas" of a pulsar's 
wind-zone -- in the language of Ruderman (1980) -- via inverse-Compton losses of the 
charges (Kundt 1998b). This interpretation is consistent with an often encountered phase shift between the radio and the hard pulses. For similar reasons, I expect all the other (incoherent) pulsar emissions, between high radio frequencies and $\gamma$-rays, to be emitted beyond the 
light cylinder.

$\bullet$ (2) How strong are \emph{pulsar winds}, measured in units of the Goldreich-Julian rate $\dot N_{GJ}:= \mu_{\parallel}\Omega^2 /ec$, ($\mu := BR^3$)? Are they leptonic? Mapped bowshocks of $\ga$ 13 pulsars indicate $\dot N$ = $\xi \dot N_{GJ}$ with $\xi \approx 10^4$, 
(Kundt 1998a). The alternative interpretation of the bowshocks -- as driven by the pulsar's outgoing Poynting flux -- is at variance with our understanding of plasma electrohydrodynamics according to which a convected (toroidal) magnetic flux would require an at least comparable 
inertia in the flux-anchoring plasma, and a strong outgoing wave would fail to sweep the blocking CSM.

Such strong winds are independently thought to be required for the high \emph{coherence} of the radio pulses, $T_b \la 10^{30 \pm 2}K$, wanting some $10^{14}$ electrons to radiate in phase (Kundt 1998a). They need not conflict with the rule that a medium does not allow wave 
propagation below its plasma frequency: A layered density profile of the (luminally!) outgoing wind would have the high plasma frequency in the high-density layers -- the antennae -- not in the low-density medium which fills the space between them.

$\bullet$ (3) How do pulsars \emph{drive} their strong winds, inferred under (2), with a \emph{neutral excess plasma} measured by $\xi  \approx 10^4$? It has been clear from the beginning of pulsar physics that gravitational forces are too weak for driving a wind, and have the 
wrong sign, so that electric forces must be at work. But there is no stationary wind solution for (radial) electric forces of uniform sign, because the star would charge up in a light-crossing time, and halt further escape; cf. P\'etri et al.~(2002). 

Instead, \emph{oscillatory discharges} are required during which negative and positive electrons are sucked out alternatingly from the polar caps, and accelerated in the form of space-charge-limited flows. Each (highly relativistic) primary electron transfers its outward momentum to $\xi 
e^{\pm}$-pairs, via $\xi$ inverse-Compton and/or curvature photons each of which, in turn, is converted into one pair via the Erber mechanism, a photon collision with the strong, transverse magnetic field (Chang 1985). Weak -- though energetic -- downward currents to the polar caps, 
by charges of the opposite sign, guarantee ample production of $e^{\pm}$-pairs which have been predicted to form a \emph{pair corona}, of temperature $T = 10^{6.5}$K, dense enough for the forming wind to draw its alternating charges therefrom, irrespective of the work function of the 
surface (Kundt \& Schaaf 1993). A derivation from first principles of these 1-d flows -- along the open magnetic field lines through the star's center (w.r.t. which the flow is symmetrical) -- is now under construction, combining the textbook treatment of a long cable with the driving, largely screened unipolar-induction voltage 
whose AC discharges satisfy a time-dependent, relativistic Child's Law with electron energies being limited by inverse-Compton and curvature-radiation losses. 

These pair coronae, of scale height $\la 10^4$cm, may explain why recent attempts at detecting a neutron star's surface chemistry (via X-rays) have resulted in perfect \emph{blackbody spectra}, (deviations $\la 10^{-2}$), in the case of more than seven nearby neutron stars 
(Burwitz et al.~2001). These coronae are thought to grow (rapidly) in column density until they get opaque to pair annihilation, of cross section $\sigma_{ann}$, at which column density their optical depth $\tau$ to photon scattering is given by  $\tau \approx \sigma _T / \sigma _{ann}$. In strong magnetic fields, pairs 
approaching each other cannot easily get rid of their angular momentum because they are trapped in the Landau ground level of gyrations, and only suffer straight-line accelerations along $\vec B$ (via mutual attraction) during flyby, with reduced radiation by about a factor of $2^{-2}$ 
(compared to the non-magnetic case); hence their capture cross section $\sigma _{ann}$ will be of the order of $\pi  r_0^2/4 \la 10^{-1} \sigma _T$. \emph{Pair coronae} are therefore expected to be \emph{opaque} to the heat radiation from a neutron star's surface: $\tau \ga 10$. Note that 
strong, pulsed pair annihilation from the polar caps of the Crab pulsar has been observed by Massaro et al.~(1991), $10^{39.9}$ annihilations per second, consistent with above estimates.

$\bullet$ (4) Strong \emph{electric currents} in the wind-zone -- compared with the Goldreich-Julian current density $j_{GJ}:= e \vec B \vec \Omega /2 \pi$ -- are in conflict with (i) a plausible driving mechanism, (ii) their closure across the polar caps, via Pedersen currents, and (iii) the 
(moderate) X-ray intensities (from the polar caps).  

$\bullet$ (5) How strong are pulsar \emph{surface magnetic fields $B_*$}? Pure dipole fields would be (i) unstable (inside a fluid star: Flowers \& Ruderman (1977), 
hence are inconsistent. During a neutron star's birth, (ii) the increasing angular velocity of the collapsing progenitor star's 
core is expected to result in a (stabilizing) \emph{toroidal bandage}, forming a `strangled dipole', or `bandaged dipole', with large, higher multipoles of odd order. Such strong higher multipoles are also required by (iii) the indispensable Erber mechanism (for pair formation), which needs high field strengths: $B_{\perp} \ge 10^{12.6}$G  $m_e c^2 / h \nu$ for the conversion of a photon of frequency $\nu$, and (iv) small curvature radii $R_c\ (\la 
R_*)$, (Kundt 1998a). This insight requires a recalculation of the emission heights of the radio pulses, so-called `radius-to-frequency mapping', which has so far only been done for dipole fields. 

However, I see no indication of surface field strengths $B_*$ above $10^{14}G$, as have been proposed for magnetars by Thompson \& Duncan (1996).

$\bullet$ (6) How do pulsars achieve their high \emph{radio brightnesses}, $T_b \la 10^{32}K$? The Lorentz-Dirac-Abraham-Laue equation of motion allows for coherent \emph{small-pitchangle synchro-curvature radiation}, whose radiative term gains importance for $Z \tau _e \approx$ ns, 
(Kundt 1998a), with a significant conversion of kinetic into radiation energy. 

$\bullet$ (7) How strongly \emph{beamed} is pulsar radio radiation? A bulk beaming fraction $f \la 0.5$ would be inconsistent with (i) the pulsar \emph{birthrate} $(\Delta t)^{-1}$, where $\Delta t = t f / N = 10^{1.3 \pm 0.3}f$ yr is the average pulsar birth interval, $t = 10^{6.4}yr$ = average 
pulsar age (determined from the distribution of $P/\dot P$), and $N$ = number of detected pulsars $\ga 10^{5.1}$, corrected for incompleteness of detection (by means of the $N$ vs distance distribution). Note that pulsars are likely the younger twins in their cradle binaries, concluded from the high occurrence rate of -- both progenitor, and neutron-star -- binaries, and from their large 
peculiar velocities. The birthrate of neutron stars is then twice the birthrate of pulsars, one in 10 f yr  in the Galaxy, unpleasantly high if f were $\la$ 0.5. 

Another argument against significant bulk beaming is (ii) the missing \emph{synchrotron nebulae} without detected central sources. The high dynamic ranges of pulsar radio emission, $\Delta I / I = 10^{2 \pm 2}$, indicate a \emph{spiky} instantaneous radiation, both in space and time, but not an 
avoidance of large spherical angles in the sky.

$\bullet$ (8) Are the \emph{msec pulsars recycled}? I don' t think so. The \emph{distribution} of pulsar periods is certainly \emph{two-humped}, but there is a clear bridge region, between the two humps -- containing more than 12 pulsars -- and there is (i) no other property that would allow to 
draw a dividing line (between `ordinary' and `recycled' ones). Recycling (ii) ignores a braking torque, (iii) conflicts with the missing (X-ray luminous!) progenitor systems, and (iv) is at variance with the predicted much higher masses of recycled pulsars ($>2 M_{\odot}$). Note that not only the rotational but also the potential (oblateness) energy grows increasingly with the spin rate. A 
stellar core collapse is required to create a msec pulsar. The msec pulsars need not be old: their spin-down ages are probably large overestimates of their true ages, by factors of $\approx 10^3$, (Kundt 1998a).  

There is a small subset of (six) \emph{anomalous X-ray pulsars} (=AXPs), of spin period $(9 \pm 3)s$, not spin-powered, but with glitches, whose ages are not known. The AXPs range at the opposite end of the 2-humped period distribution (to the ms pulsars). Were they born slow?

$\bullet$ (9) Why are pulsar clocks \emph{noisy}, with both discrete `glitches', and continuous spin-down noise? In (Kundt 1998a) I have shown that both inaccuracies can be understood as due to a mild \emph{super-rotation}, $\Delta \Omega / \Omega \la 10^{-3}$, of their neutral, superfluid 
component, of relative moment of inertia $\Delta I / I \la 0.1$, which couples to the normal component in a stochastic way. It is possible that the gradual decay of the higher (than dipole) magnetic multipole moments yields an independent contribution to a pulsar's noisiness.

$\bullet$ (10) Can pulsars act as \emph{jet sources}? The recent, repeated X-ray maps by the CHANDRA satellite, of the near surroundings of the 4 pulsars: Crab, Vela, B1509-58 (in MSH 15-5\emph{2}, = G320.4-1.2), and  B1951+32 (in CTB 80), show a wealth of time-varying detail involving (i) a large number 
of expanding (segments of) ellipses, (ii) symmetric arrays of bright spots, and (iii) two fuzzy, antipodal, jet-like protrusions along the symmetry axis; all this out to distances of $\la 2$ lyr from the pulsar. What goes on?
  
The outer ellipses have been interpreted, for some 25 yr, as due to enhanced radiation from the (equatorial) inner edge of the pulsar nebula, at the injection shock, where the pulsar wind is decelerated -- from supersonic to subsonic -- as it joins the outward motion of earlier generations 
of injected relativistic $e^{\pm}$-pairs. New are in particular the jets, but also the bright spots, and an inner ring. 

The jets look like of \emph{Eilek-type B}, i.e. like relativistic jets whose (decelerated) heads move subsonically w.r.t. their CSM, so that they get defocussed, and lack hot-spots at their outer ends. Jet-sources are thought to evolve from (edge-brightening) type A to (edge-dimming) type B, as a 
consequence of a decreasing ram pressure and (perhaps) an increasing ambient sound speed with increasing jet length (Kundt 2001). In the present case (of pulsars), the jets form an energetically minor constituent inside a relativistically hot environment, hence look weak already at their young ages. They look very different from the strong jets of, say, SS 433, or of some of the black-hole 
candidates, which latter I understand as neutron stars surrounded by heavy accretion disks. 

These new \emph{pulsar jets} tell me that our earlier understanding of pulsars has been overidealized, as isolated neutron stars inside clear skies which have swept clean a cavity around them, of radius some lightyear. Pulsar winds are too light to achieve this. Rayleigh-Taylor 
instabilities allow debris from the SN explosion to hang around and/or fall back, to halt the pulsar wind near its polar caps, confine it, and get it re-started, as a feeble twin jet. The transiently trapped, highly relativistic pairs escape supersonically through a self-rammed, narrow channel, and blow lobes whose intersections with the nozzle are seen glowing at X-rays, as rings and bright 
spots, the latter at local density enhancements of the debris. X-ray movies which have been prepared from successive maps are reported to show perplexing variability: We see locally heated matter pushed around by new generations of escaping pulsar-wind leptons.

\section{Eight further Alternatives related (more generally) to Neutron Stars} 

$\bullet$ (11) Do (essentially) all \emph{supernovae} give birth to a neutron star? Are they thick-walled explosions, i.e. splinter bombs, implying that their remnants glow via the collisional heating of their CSM, as maintained, e.g., in (Kundt 2001)? Note the particular morphology of the Vela 
remnant, with its large protrusions seen at X-rays (Aschenbach et al.~1995). 

$\bullet$ (12) Must supernova remnants be distinguished from \emph{pulsar nebulae} -- like CTB 80, or G 5.3-1.0, the `Bird' -- the latter illuminated by a pulsar wind (Kundt \& Chang 1992)? Of particular concern are the \emph{soft $\gamma$-ray repeaters} (= SGRs), whose distances can be 
largely overestimated, and ages largely underestimated if assumed inside SNRs.

$\bullet$ (13) Can neutron stars act as \emph{ejectors} -- in distinction to pulsars, or (accreting) binary X-ray sources -- like SS 433, and create pair-plasma bubbles which escape as \emph{jets} (Kundt 1998a; 2001)? I.e. neutron stars may behave in many different ways: like pulsars, 
accretors, ejectors, and `dead neutron stars', among them perhaps the cosmic-ray boosters, the black-hole candidates, and the $\gamma$-ray bursters. Note that SS 433 has recently shown that (even) jet formation need not be without losses: Blundell et al.~(2001) find radio emission in the 
orbit plane, at distances $\ga 10^2$ AU. Those who do not believe in in-situ acceleration must conclude that spillover relativistic pairs are convected outward by an equatorial wind.

$\bullet$ (14) Are the \emph{cosmic rays} generated by the corotating magnetospheres of neutron stars, via \emph{relativistic slingshot} ejections (Kundt 2001)? Note that 20 \% of all CRs above $10^{20}$eV are \emph{repeaters}, i.e. come from the same direction (within $1^0$), asking for 
nearby compact sources, and that the CRs are hydrogen- and helium-deficient.

$\bullet$ (15) Are the (bright) \emph{supersoft X-ray sources} -- with energies between 20 and 60 eV -- emitted by forming massive accretion disks (between 1 and 20 $M_{\odot}$, typically 5 $M_{\odot}$), rather than by white dwarfs (Kundt 1996)? I think so. The record in mass is 
presently held by GRS 1915+105, (Greiner et al.~2001).

$\bullet$ (16) How to explain the \emph{super-Eddington sources}, mostly (agreed) neutron-star binaries with X-ray powers $\la 10^{41}$erg/s? Are they due to `blade' accretion from forming massive disks, i.e. due to magnetically confined, clumpy accretion in the orbit plane 
(Kundt 1996)?

$\bullet$ (17) Are the \emph{black-hole candidates} neutron-star binaries with massive accretion disks, filling the gap between the low-mass and high-mass X-ray binaries (Kundt 2001)? 

$\bullet$ (18) Are the \emph{$\gamma$-ray bursts} emitted by spasmodically accreting old, nearby (Galactic) neutron stars, via clumpy accretion, as is signaled by the ($\ga$ 6) repeaters? The SGRs are the nearest among them, from which we receive both the rare, ordinary GRBs and, in 
addition, the frequent softer and vastly weaker repetitions. Their afterglows -- whose relativistic redshifts are reminiscent of the transverse Doppler shifts of SS 433 --  would be \emph{light echos} from their transient hadronic winds as well as from their CSM.

\begin{acknowledgements}
There were times in modern sciences when one could use a significant fraction of one's working 
hours for doing research, with no extra education required in how to please a group of malicious 
PCs; these times now belong to the past. This communication owes its existence to my younger 
colleagues Robindro Dutta-Roy, J\"urgen P\"apke, and Gernot Thuma for whose repeated help I 
thank them warmly.

Concerning the science part, my thanks go to Bernd Aschenbach, Joachim Tr\"umper, and to Gernot Thuma 
for in-depth, undoubtedly honest, and always ready discussions respectively. 
\end{acknowledgements}


\clearpage


\begin{thebibliography}{}
\bibitem{AE95} Aschenbach B., Egger R., Tr\"umper J., 1995, Nature 373, 587-590
\bibitem{BMMPR01} Blundell K.M., Mioduszewski A.J., Muxlow T.W.B., Podsiadlowski P., Rupen M.P., 2001, ApJ 562, L79-L82
\bibitem{BZ01} Burwitz V., Zavlin V.E., Neuh\"auser R., Predehl P., Tr\"umper J., Brinkman A.C., 2001, A\&A 379, L35-L38
\bibitem{C95} Chang H.-K., 1995, A\&A 301, 456-462
\bibitem{FR77} Flowers E., Ruderman M.A., 1977, ApJ 215, 302-310
\bibitem{GC01} Greiner J., Cuby J.G., McCaughrean M.J., 2001, Nature 414, 522-524
\bibitem{K96} Kundt W., 1996, in Supersoft X-Ray Sources, ed. J. Greiner, Lecture Notes in Phys. 472, 45-50
\bibitem{K98a} Kundt W., 1998a, Fundamentals of Cosmic Phys. 20, 1-119
\bibitem{K98b} Kundt W., 1998b, Mem. Soc. Astron. Ital. 69, 911-917
\bibitem{K01} Kundt W., 2001, Astrophysics - A Primer, Springer
\bibitem{KC92} Kundt W., Chang H.-K., 1992, Ap\&SS 193, L145-L154
\bibitem{KK80} Kundt W., Krotscheck E., 1980, A\&A 83, 1-21
\bibitem{KS93} Kundt W., Schaaf R., 1993, Ap\&SS 200, 251-270
\bibitem{MM91} Massaro E., Matt G., Salvati M., Costa E., Mandrou P., Niel M., Olive J.F., Mineo T., Sacco B., Scarsi L., Gerardi G., Agrinier B., Barouch E., Comte R., Parlier B., Masnou J.L., 1991, ApJ 376, L11-L14
\bibitem{PH02} P\'etri J., Heyvaerts J., Bonazzola S., 2002, A\&A 384, 414-432
\bibitem{R80} Ruderman M., 1980, in Ninth Texas Symposium on Relativistic Astrophysics, eds. J. Ehlers, J. Perry, \& M. Walker, Ann. New York Acad. Sci. 336, 409-428
\bibitem{TD96} Thompson C., Duncan R., 1996, ApJ 473, 322-342
\end{thebibliography}
\end{document}